\documentclass[ aps,%
floatfix,%
final,%
notitlepage,%
oneside,%
onecolumn,%
nobibnotes,%
nofootinbib,%
superscriptaddress,%
showpacs,%
]%
{revtex4}

\usepackage{epsfig}
\usepackage{amsfonts}
\usepackage{amsmath}
\usepackage{epsfig}
\usepackage{graphics}

\newcommand{\beq}{\begin{eqnarray}}\newcommand{\eeq}{\end{eqnarray}}
\newcommand{\beqa}{\begin{eqnarray*}}\newcommand{\eeqa}{\end{eqnarray*}}

\begin{document}

\title{Properties of the quark-antiquark-gluon Fock state in the $\eta_c$ meson}
\author{V.V. Braguta}
\email{braguta@mail.ru}
\affiliation{Institute for High Energy Physics, Protvino, Russia}
\affiliation{Institute for Theoretical and Experimental Physics, Moscow, Russia}

\begin{abstract}
In this paper the twist-3 distribution amplitude of the quark-antiquark-gluon Fock state in the 
$\eta_c$ meson is studied. To calculate the moments of this distribution amplitude 
QCD sum rules is applied. Using the results of the calculation the model of this
distribution amplitude is built. In addition NRQCD matrix elements which 
determine the properties of the quark-antiquark-gluon Fock state are determined. 
In particular, the probability amplitude to find the quark-antiquark pair in the color octet 
state, the mean gluon energy and the fraction of momentum carried by gluon, the relative velocity 
of the color-octet quark-antiquark-pair are calculated.
\end{abstract}


\maketitle

\newcommand{\ins}[1]{\underline{#1}}
\newcommand{\subs}[2]{\underline{#2}}
\vspace*{-1.cm}
\section{Introduction.}

Theoretical description of hard exclusive processes  is based on 
the factorization theorem \cite{Lepage:1980fj, Chernyak:1983ej}. Within this theorem the amplitude of hard exclusive process
can be separated into two parts. The first part is partons production at very small 
distances, which can be treated within perturbative QCD. The second part is 
hardronization of the partons at larger distances. This part contains  information about 
nonperturbative dynamic of the strong interactions and it can be 
parameterized by process independent distribution amplitudes (DA), which can be considered as wave functions of hadrons
at light-like separation between the partons. 

Recently, two-particle charmonia DAs  have become an object of intensive 
study \cite{Bodwin:2006dm, Ma:2006hc, Braguta:2006wr, Braguta:2007fh, Braguta:2007tq, Choi:2007ze, Feldmann:2007id, 
Bell:2008er, Braguta:2008qe, Hwang:2008qi, Hwang:2009cu, Braguta:2009ej}.
Research in this direction allowed one to build models for two-particle charmonia DAs and 
to carry out the study of some interesting exclusive processes with charmonia production 
(see, for instance, \cite{Braguta:2008tg, Braguta:2009df, Braguta:2009xu}). 

The first aim of this paper is to study the properties of the twist-3 quark-antiquark-gluon DA of the 
$\eta_c$ meson within QCD sum rules and to build the model of this function which can be used in the 
calculation of different production processes. There are a lot of papers
(see, for instance, \cite{Chernyak:1983ej, Braun:1988qv, Ball:1998je}) devoted to
the study  of the quark-antiquark-gluon Fock state DAs of light pseudoscalar mesons. 
However, until now there are no papers devoted to the study of similar DAs for charmonia. 

Another description of charmonia production processes, which is called nonrelativistic QCD(NRQCD), 
is based on the expansion of matrix elements in powers of the relative velocity of quark-antiquark pair inside 
charmonia mesons \cite{Bodwin:1994jh}. NRQCD matrix elements play crucial role in this approach. 
These matrix elements parameterize charmonia structure. For instance, 
there is the  matrix element which determines the probability amplitude to find quark-antiquark 
pair in the color-octet state in the $\eta_c$ meson. Another example is the matrix element 
which determines the gluon energy in the $\eta_c$ meson. The second aim 
of this paper is to determine some of these matrix elements from 
the quark-antiquark-gluon DA of the $\eta_c$ meson.

This paper is organized as follows. In the next section the definition of the 
twist-3 quark-antiquark-gluon DA  will be given. 
The parameters of this DA will be calculated in the third section. 
The forth section is devoted to the calculation of the NRQCD matrix elements 
which parameterize the quark-antiquark-gluon Fock state in the $\eta_c$ meson. 
In section V the model of the DA under consideration will be proposed. 
In the last section the results of this paper will be summarized.

\section{Definition.}

The twist-3 distribution amplitude (DA) of the quark-antiquark-gluon Fock state in 
the $\eta_c$ meson can be defined as 
follows \cite{Braun:1988qv, Ball:1998je}
\beq
\label{f:eq}
\langle \eta_c (p) | \bar q (z) [z,vz] \sigma_{\mu \nu} \gamma_5 g G_{\alpha \beta} (v z) [vz,-z] q(-z) | 0 \rangle_Q = 
i f_{3 \eta} \biggl (
p_{\alpha} p_{\mu} g^{\perp}_{\nu \beta} - p_{\alpha} p_{\nu} g^{\perp}_{\mu \beta} - p_{\beta} p_{\mu} g^{\perp}_{ \nu \alpha} + 
p_{\beta} p_{\nu} g^{\perp}_{\mu \alpha}
 \biggr ) \times \\ \nonumber 
\times \int d x_1 d x_2 d x_3 \delta (x_1+x_2+x_3-1) e^{i (pz)(x_1-x_2+v x_3) } V_Q(x_1, x_2, x_3),
\eeq
where $V_Q(x_1,x_2,x_3)$ is the twist-3 DA of the quark-antiquark-gluon Fock state, 
$x_1, x_2, x_3$ are the fractions of momentum of the $\eta_c$ meson carried by the quark, antiquark and gluon correspondingly,
\beq
g^{\perp}_{\mu \nu} = g_{\mu \nu} - \frac {p_{\mu} z_{\nu} + p_{\nu} z_{\mu}} {pz},\nonumber \\ \nonumber
[z_1, z_2]=P \exp {\biggl (  i g \int_{z_1}^{z_2} dz^{\mu} A_{\mu}  \biggr )}. 
\eeq
Below it will be assumed that the function $V_Q(x_1, x_2, x_3)$ is normalized as follows
\beq
\int d x_1 d x_2 d x_3 \delta (x_1+x_2+x_3-1)  V_Q(x_1, x_2, x_3) =1
\label{norm}
\eeq
The DA $V_Q(x_1, x_2, x_3)$ is defined at the renormalization scale $Q$. 
The renormalization properties of three-particles DAs are rather complicated and they will not be discussed in this paper. 
For this reason below the subscript $Q$ will be omitted. It is worth to note that if the energy 
scale tends to infinity $Q \to \infty$, the DA $V_Q(x_1,x_2, x_3)$ tends to it's asymptotic 
form
\beq
V_{as}(x_1, x_2, x_3) = 360 x_1 x_2 x_3^2.
\eeq
Because of nonrelativistic nature of heavy quarkonia, one can expect that  real DA is 
very far from its asymptotic form. 

Another very useful property of the DA is that the function $V(x_1,x_2,x_3)$ is symmetric under the replacement 
$x_1 \leftrightarrow x_2$:
\beq
V(x_1,x_2,x_3)=V(x_2,x_1,x_3).
\label{sym}
\eeq

Below we will also need the relation between the moments of the DA $V(x_1, x_2, x_3)$ and QCD matrix elements
\beq
2 i f_{3 \eta} (pz)^{n_1+n_2+n_3+2} \langle x_1^{n_1} x_2^{n_2} x_3^{n_3}\rangle =  \langle \eta_c(p) | \bigl \{ \bar q (i z \overset {\leftarrow} {D})^{n_1} \bigr \}
\sigma_{\mu \lambda} \gamma_5 \bigl \{  g (i z \overset {\rightarrow} {D})^{n_3} G_{\mu \nu} \bigr \} \bigl \{ (i z \overset {\rightarrow} {D})^{n_2} q \bigr \} |0 \rangle z^{\lambda} z^{\nu}.
\label{matrix_el}
\eeq
The properties of the DA $V(x_1,x_2,x_3)$ can be parameterized by the moments $\langle x_1^{n_1} x_2^{n_2} x_3^{n_3}\rangle$. 
In this paper the following moments will be calculated $\langle x_3 \rangle, \langle x_3^2 \rangle, \langle (x_1-x_2)^2 \rangle$. 
The other the first and the second moments can be expressed through these ones.

\section{The moments of the distribution amplitude $V(x_1,x_2,x_3)$ from QCD sum rules.}

QCD sum rules \cite{Shifman:1978bx, Shifman:1978by} proved to be very effective and universal tool in the 
determination of different  matrix elements. In particular, QCD sum rules can be applied 
to study the properties of  DAs.  There are a lot of papers
( see, for instance, \cite{Chernyak:1983ej, Braun:1988qv, Ball:1998je} ) where QCD sum rules was applied 
to study the parameters of the twist-3 DAs of light pseudoscalar mesons. 
However, there are no papers devoted to the study of  heavy quarkonia DAs of the quark-antiquark-gluon 
Fock state. 

In this section QCD sum rules will be applied to calculate the moments of the function $V(x_1,x_2,x_3)$. 
To do this let us consider the correlator 
\beq
\Pi_{n_1, n_2, n_3}(q,z)=i \int d^4 x e^{iqx} \langle 0|  T J^+_{n_1, n_2, n_3}(x,z) J_{0,0,0}(0,z)   |0 \rangle= (zq)^{n_1+n_2+n_3+4} \Pi(q^2),
\label{cor1}
\eeq
where the current $J_{n_1, n_2, n_3}(x,z)$ is defined as follows
\beq
J_{n_1, n_2, n_3}(x,z)=\bigl \{ \bar q (i z \overset {\leftarrow} {D})^{n_1} \bigr \}
\sigma_{\mu \lambda} \gamma_5 \bigl \{  g (i z \overset {\rightarrow} {D})^{n_3} G_{\mu \nu} \bigr \} \bigl \{ (i z \overset {\rightarrow} {D})^{n_2} q \bigr \} 
z^{\lambda} z^{\nu}.
\eeq
Applying standard procedure which will not be described here one gets QCD sum rules for 
correlator (\ref{cor1})
\beq
4 f_{3 \eta}^2 \frac {\langle x_1^{n_1} x_2^{n_2} x_3^{n_3}\rangle} {(m_{\eta_c}^2+Q^2)^{m+1}} = \frac 1 {\pi}
\int_{4 m_c^2}^{s_0} ds \frac {\mbox {Im} \Pi_{\rm pert}( s)} {(s+Q^2)^{m+1}} + \frac 1 {m!} \biggl ( - \frac {d} {d Q^2}  \biggr )^{m} \Pi_{\rm npert} (Q^2),
\label{sm1}
\eeq
where $Q^2=-q^2$, $\Pi_{\rm pert}( s), \Pi_{\rm npert}( s)$ are the perturbative and nonperturbative contributions to correlator (\ref{cor1}). 
The nonpertubative part of the correlator parameterizes the contribution of the vacuum condensates, which at the leading order 
approximation used in this paper is given by the gluon vacuum condensate $\langle \alpha_s G^2 \rangle$. 
The expressions for the $\mbox{Im} \Pi_{\rm pert}( s), \Pi_{\rm npert}( s)$ are rather lengthy and very complicated. 
For this reason they will not be shown here. 

In the original paper \cite{Shifman:1978by} 
the method QCD sum rules was applied at $Q^2=0$. However, as was shown in paper \cite{Reinders:1984sr}
there is a large contribution of higher dimensional operators at $Q^2=0$ which grows rapidly with $m$. 
To suppress this contribution in this paper sum rules (\ref{sm1}) will be applied at $Q^2=4 m_c^2$. 

In the numerical analysis of QCD sum rules the values of the parameters $m_c$ and $\langle \alpha_s G^2 \rangle$ will be 
taken from paper: \cite{Reinders:1984sr}
\beq
m_c = 1.24 \pm 0.02 ~\mbox {GeV}, ~~ \langle \frac {\alpha_s} {\pi} G^2 \rangle = 0.012 \pm 30 \% ~\mbox {GeV}^4.
\label{param}
\eeq
The value of the threshold $s_0$ is varied within the 
interval $\sqrt {s_0} \in (3.7 \pm 0.2)$ GeV. The value of the strong coupling constant is taken from paper \cite{Baikov:2008jh}
\beq
\alpha_s(m_{\tau})=0.332 \pm 0.015,
\eeq
and then it is evolved to the scale $\mu=m_c$ at two loops. In the calculation 
the uncertainty of the results due to the variation of the strong coupling 
constant $\alpha_s$ is not very important. For this reason it will be disregarded. 

The application of sum rules (\ref{sm1}) to the calculation of the constant $f_{3\eta}$
gives the following result
\beq
f_{3\eta}^2=(12 \pm 3 \pm 2 \pm 1)\cdot 10^{-6}~ \mbox{GeV}^4,
\label{f3eta}
\eeq
where the second, the third and the forth errors are due to the uncertainties in the values 
of the parameters $m_c$, $\langle \alpha_s G^2 \rangle$ (\ref{param}) and $s_0$ correspondingly. 

Further let us proceed to the calculation of the moments $\langle x_1^{n_1} x_2^{n_2} x_3^{n_3} \rangle$. 
The result of the calculation of the moment $\langle x_3 \rangle$ is
\beq
\langle x_3 \rangle =0.22 \pm 0.01 \pm 0.01 \pm 0.02.
\label{x3eta}
\eeq
The sources of errors are the same as  in (\ref{f3eta}). Evidently, 
due to the symmetry relation (\ref{sym}) the other first 
 moments can be related to the $\langle x_3 \rangle$ as follows $\langle x_1 \rangle=\langle x_2 \rangle= \langle (1-x_3)/2 \rangle$. 

It should be noted that  moment (\ref{x3eta}) has simple and very important physical meaning. 
The $\langle x_3 \rangle$ measures the fraction of momentum carried by gluon in the 
quark-antiquark-gluon Fock state of the $\eta_c$ meson. Value (\ref{x3eta}) tells us that 
the gluon carries $\sim 20 \%$ of the total momentum. 

Another moments which also have important physical meaning are $\langle (x_1-x_2)^2 \rangle, \langle x_3^2 \rangle$. 
The calculation gives
\beq
\langle (x_1-x_2)^2 \rangle = 0.021 \pm 0.002 \pm 0.003 \pm 0.001 \nonumber \\
\langle x_3^2 \rangle=0.055 \pm 0.005 \pm 0.006 \pm 0.004. 
\label{x32eta}
\eeq
The sources of errors are the same as  in (\ref{f3eta}).
It is easy to show that all moments of the first and the second order can be expressed through moments 
(\ref{x3eta}), (\ref{x32eta}). 

It should be noted here that matrix elements (\ref{f3eta}), (\ref{x3eta}), (\ref{x32eta}) are
scale dependent quantities. The virtuality of the propagators in the calculation 
of the $\mbox{Im} \Pi_{\rm pert}( s), \Pi_{\rm npert}( s)$ is of order of $\sim m_c$. 
So, results (\ref{f3eta}), (\ref{x3eta}), (\ref{x32eta}) are defined at the same scale. 

\section{NRQCD matrix elements and properties of the quark-antiquark-gluon Fock state in the $\eta_c$ meson.}

As was noted in the introduction, the DA $V(x_1,x_2, x_3)$ contains the information 
about properties of the quark-antiquark-gluon Fock state in the $\eta_c$ meson. According to formula (\ref{matrix_el}) 
this properties can be parameterized by some rather complicated QCD operators. To make the 
result of this paper more transparent, one can expand these operators in the relative velocity  of the 
quark-antiquark pair in the $\eta_c$ meson, as was done in paper \cite{Braguta:2006wr}. 
Following the approach proposed in this paper, at the leading order approximation one gets the relations
\beq
\langle \eta_c | \psi^+ \sigma_j \bigl ( gH_j \bigr ) \chi   | 0 \rangle = 3 f_{3\eta} M_{\eta}^2, 
\label{f1} \\
\langle \eta_c | \psi^+ \sigma_j \bigl (    (i \overset {\rightarrow} \nabla_0) gH_j \bigr ) \chi   | 0 \rangle = 3 f_{3\eta} M_{\eta}^3 \langle x_3 \rangle, 
\label{f2} \\
\langle \eta_c | \psi^+ \sigma_j \bigl (   ( -3 \overset {\rightarrow}\nabla_0^2 - \overset {\rightarrow} \nabla_k^2 ) gH_j \bigr ) \chi   | 0 \rangle = 9 f_{3\eta} M_{\eta}^4 \langle x_3^2 \rangle, 
\label{f3} \\
\langle \eta_c | \psi^+ \sigma_j \bigl (   (  -\overset {\leftarrow}\nabla_k^2 gH_j  - gH_j \overset {\rightarrow}\nabla_k^2 +2 
\overset {\leftarrow}\nabla_k gH_j \overset {\rightarrow}\nabla_k )  \bigr ) \chi   | 0 \rangle = 9 f_{3\eta} M_{\eta}^4 \langle (x_1-x_2)^2 \rangle. 
\label{f4} 
\eeq
where $\psi^+$ and $\chi$ are Pauli spinor fields that create a quark and an antiquark respectively, $\vec {\sigma}$
is the Pauli matrices, $\vec H$ is a chromomagnetic field, $j,k$ are spatial indexes. Note that in (\ref{f1})-(\ref{f4}) summation over the repeated 
indexes is assumed. 

From formulas (\ref{f1})-(\ref{f4}) one sees that the quark-antiquark pair is in the $S$-wave spin-triplet 
color-octet state. The role of the color-octet state in the $\eta_c$ meson is determined 
by the matrix element
\beq
\langle \eta_c | \psi^+   g \bigl ( \vec {\sigma} \vec {H} \bigr ) \chi   | 0 \rangle = 3 f_{3\eta} M_{\eta}^2=(92 \pm 34) \cdot 10^{-3}~ \mbox{GeV}^4.
\eeq
Note that in the last formula the uncertainty due to the 
higher order terms in the relative velocity expansion have been taken into account. 

Strictly speaking, the role of the color-octet state in the $\eta_c$ meson strongly depends 
on the process. However, one can estimate the probability amplitude to find the 
color-octet state as follows
\beq
\langle \bar q q G | \eta_c \rangle \sim \frac {\langle \eta_c | \psi^+ \sigma_j \bigl ( gH_j \bigr ) \chi   | 0 \rangle} 
{m_c^2 \langle \eta_c | \psi^+  \chi   | 0 \rangle} \sim 0.04.
\label{prob1}
\eeq
One sees that this value is very small. In paper \cite{Braaten:1998au} the ratio $\langle \bar q q G | \eta_c \rangle$
was estimated as $\sim v^{7/2} \sim 0.06 - 0.09$ for $v^2 \sim 0.20 - 0.25$, what is in reasonable agreement 
with result (\ref{prob1}).

Now one can estimate the average fraction of momentum carried by gluon in the $\eta_c$ meson. To do this 
one can apply the standard quantum mechanical formula
\beq
\langle x_g \rangle_{\eta}=\sum_{f} p_f \times \langle x_g \rangle_{f}, 
\label{prob}
\eeq
where the sum is taken over all Fock states $f$ in the $\eta_c$ meson, $p_f$ is 
the probability to find the Fock state $f$, $\langle x_3 \rangle_{f}$
is the fraction of momentum carried by gluon in the Fock state $f$. Assuming
that the dominant contribution to formula (\ref{prob}) is given by the 
quark-antiquark-gluon Fock state one gets $\langle x_g \rangle_{\eta} \sim 3 \cdot 10^{-4}$.

Further let us pay attention to formula (\ref{f2}). This formula allows us to 
calculate the mean energy of the gluon in the $\eta_c$ meson
\beq
E_g= \frac {\langle \eta_c | \psi^+ \sigma_j \bigl (    (i \overset {\rightarrow} \nabla_0) gH_j \bigr ) \chi   | 0 \rangle } 
{ \langle \eta_c | \psi^+ \sigma_j \bigl ( gH_j \bigr ) \chi   | 0 \rangle }=M_{\eta} \langle x_3 \rangle=(0.66 \pm 0.18)~ \mbox{GeV}.
\label{eg}
\eeq
One can also estimate the gluon energy from formula (\ref{f3})
\beq
E_g \sim \sqrt { \frac {\langle \eta_c | \psi^+ \sigma_j 
\bigl (   ( -3 \overset {\rightarrow}\nabla_0^2 - \overset {\rightarrow} \nabla_k^2 ) gH_j \bigr ) \chi   | 0 \rangle }
{ 4 \langle \eta_c | \psi^+ \sigma_j \bigl ( gH_j \bigr ) \chi   | 0 \rangle } }= M_{\eta} \sqrt { \frac 3 4  \langle x_3^2 \rangle } =0.61 ~ \mbox{GeV}.
\eeq
This value is in a good agreement with (\ref{eg}).

In the last excise the relative momentum ${\bf p_r}$ of the color-octet quark-antiquark pair will be determined. 
To do this let us apply formula (\ref{f4})
\beq
|{\bf p_r}| \sim \frac 1 {2 (1-\langle x_3 \rangle)} \sqrt {
\frac {\langle \eta_c | \psi^+ \sigma_j \bigl (   (  -\overset {\leftarrow}\nabla_k^2 gH_j  - gH_j \overset {\rightarrow}\nabla_k^2 +2 
\overset {\leftarrow}\nabla_k gH_j \overset {\rightarrow}\nabla_k )  \bigr ) \chi   | 0 \rangle}
{\langle \eta_c | \psi^+ \sigma_j \bigl ( gH_j \bigr ) \chi   | 0 \rangle}
} = \frac 1 2 \sqrt { 3 \frac { \langle (x_1-x_2)^2 \rangle} {(1- \langle x_3 \rangle)^2}  } M_{\eta} = 0.49 ~ \mbox{GeV},
\eeq
It is important to note that the relative velocity of the color-octet quark-antiquark pair
($v_o^2 \sim 3 \langle (x_1-x_2)^2 \rangle / (1-\langle x_3 \rangle)^2  \sim$ 0.11) is two times smaller than 
that for the color-singlet quark-antiquark state($v_s^2 \sim 0.21$ \cite{Braguta:2006wr}).
What is expected result since the gluon carries some fraction of  momentum and  energy of the $\eta_c$ meson. 

\section{The model of the distribution amplitude $V(x_1, x_2, x_3)$.}

To build the model of the DA  $V(x_1, x_2, x_3)$ one can recall that the models of 
the leading twist charmonia distribution amplitudes proposed in papers 
\cite{Braguta:2006wr, Braguta:2007fh, Braguta:2008qe} can be written in the 
following form
\beq
\phi \sim \phi_{as} \cdot \exp { \biggl ( - \frac { M^2_{c \bar c}} {M^2}  \biggr )},
\eeq
where $\phi_{as}$ is the asymptotic form of the distribution amplitude $\phi$, 
$M^2_{c \bar c}$ is the invariant mass of the $c \bar c$ system expressed in terms 
of kinematic variables, $M^2$ is the characteristic invariant mass of the 
$c \bar c$ system. 

In the case of the quark-antiquark-gluon Fock state the $\eta_c$ meson can be divided 
into two subsystems: the quark-antiquark pair in the color-octet state and the color-octet pair with 
the gluon. If one introduces the following variable 
\beq
x_1=xy,~~~
x_2=(1-x)y,~~~
x_3=1-y,
\eeq
the invariant masses of the corresponding subsystems can be written in the form
\beq
M^2_{c \bar c} = m_c^2 \frac {1} {x(1-x)} = m_c^2 \frac {(1-x_3)^2} {x_1 x_2},~~~ M^2_{ \{ c \bar c  \}g}=m_c^2 \frac {1} {xy(1-x)}=m_c^2 \frac {(1-x_3)} {x_1x_2}.
\eeq
To build the model of the DA $V(x_1, x_2, x_3)$ one should introduce the 
exponent factors for the both subsystems. Thus one has
\beq
V(x_1, x_2, x_3) = c(\beta_1, \beta_2) x_1 x_2 x_3^2 \cdot \exp { \biggl ( - \beta_1 \frac {(1-x_3)^2} {x_1 x_2}  \biggr )} 
\cdot \exp { \biggl ( - \beta_2 \frac {(1-x_3)} {x_1x_2}  \biggr )},
\label{model}
\eeq
where the constant $c(\beta_1, \beta_2)$ can be determined from normalization condition (\ref{norm}). 
The values of the parameters $\beta_1, \beta_2$ can be fixed from the requirement that 
the values of the moments $\langle x_3 \rangle, \langle (x_1-x_2)^2 \rangle$ for model (\ref{model}) 
must coincide with the values (\ref{x3eta}), (\ref{x32eta}). Thus one gets 
$\beta_1=1.25_{-0.45}^{+0.65} , \beta_2=1.15 \pm 0.25$.  Note that model (\ref{model}) is defined at the scale $\mu \sim m_c$. 

One can assume that it is sufficient to introduce only one exponent with the total 
invariant mass of the quark-antiquark-gluon system in formula (\ref{model}). However, 
the calculation shows that in this case it is not possible to reproduce 
values (\ref{x3eta}), (\ref{x32eta}) of the moments $\langle x_3 \rangle, \langle (x_1-x_2)^2 \rangle$  simultaneously. Physically, 
this means that although  energy scales of the two subsystems introduced above are
of the same order they  are not equal to each other.

\section{Conclusion}

In this paper the properties of the quark-antiquark-gluon Fock state 
in the $\eta_c$ meson are studied. This properties can be parameterized by the moments of the 
twist-3 distribution amplitude (\ref{f:eq}) and the constant $f_{3 \eta}$. To calculate 
the first, the second order moments and the constant $f_{3 \eta}$ QCD sum rules was applied. 
The result of this calculation allowed one to build the model of the DA which can be used 
in the study of different production processes. 

To make the results of the calculation more transparent, QCD operators which
determine the moments of the distribution amplitude are expanded
in relative velocity of the quark-antiquark pair in the $\eta_c$ meson. 
In particular, it was shown 
that the quark-antiquark pair is in the $S$-wave spin-triplet color-octet state. 
The probability amplitude 
to find the quark-antiquark pair in the color-octet state is rather small $\sim 0.04$. The energy 
of the gluon in the quark-antiquark-gluon Fock state is $\sim 600$ MeV. The relative velocity of the color-octet 
quark-antiquark pair is $\sim 0.11$ what is much smaller than that in the color-singlet 
quark-antiquark pair.

\begin{acknowledgments}
I would like to thank A.K. Likhoded and A.V. Luchinsky for useful discussion.
This investigation has been supported by grants RFBR 10-02-00061-a, RFBR 11-02-00015-a, RFBR 11-02-01227-a and
by the Federal Special-Purpose Program
’Cadres’ of the Russian Ministry of Science and Education.

\end{acknowledgments}

\end{document}